# The WARP Reactor Concept

M. G. Anderson[1], J. K. Walters[1], E. M. Anaya[1], D. A. Max[2], W. A. Stygar[1] and A. J. Link[1]

[1]*Lawrence Livermore National Laboratory, Livermore, California 94550, USA*
[2]*Mission Support and Test Services, Livermore, California 94550, USA*

(Dated: 08 June 2023)

The WARP Reactor Concept promises orders of magnitude increase of intense ion beam energies and respective radiation yields at a fraction of the size and cost over existing z-pinch class accelerators allowing the economically viable study of new Relativistic High Energy Density Physics regimes for probing the intersection between General Relativity and Quantum Field Theory along with game-changing direct applications from rep-rated Magnetized Liner Inertial Fusion devices for energy production and advanced propulsion to multi-pulse compact flash x-ray/neutron radiography sources for assessing our Nation's aging nuclear weapons stockpile. An overview of the WARP Reactor Concept is presented.

## I. INTRODUCTION

The Wave Accelerated Ring Pinch or "WARP" Reactor [1,2] (patent pending), iso-view shown in Figure 1, is expected to solve key issues ranging from our present energy dependence on finite fossil fuels and its associated climate impact to our aging nuclear weapons stockpile. The WARP Reactor promises orders of magnitude increase of ultra-intense ion beam energies and respective high radiation yields at a fraction of the size and cost over other z-pinch class accelerators [3-5] allowing the economically viable and environmental friendly study of new Relativistic High Energy Density (RHED) Physics regimes for probing the intersection between General Relativity and Quantum Field Theory (i.e. Warp/Unruh/Casimir effects) [6-24] along with game-changing direct applications from rep-rated Magnetized Liner Inertial Fusion (MagLIF) devices for energy production and advanced propulsion to multi-pulse compact flash x-ray/neutron radiography sources [3-5,25] for assessing our Nation's aging nuclear weapons stockpile.

## II. NOVELTY

The WARP Reactor concept is a novel, modular and compact pulsed power-driven radiation source intended for nuclear fusion energy production, advanced propulsion, accessing new RHED physics regimes and flash radiography/interrogation techniques. WARP utilizes state-of-the-art pulsed power modules to drive its "WARP Core" which consists primarily of two Dense Plasma Focuses (DPFs) [26,27] and two Ion Ring Marx Generators (IRMGs) fired directly at one another. The WARP Reactor's dramatic performance boost is achieved with the use of a novel WARP Core which injects two tubular dense plasma and ion beams from opposite ends of a double-barreled DPF head with embedded IRMG-driven reflex triodes [28] and through magnetic cusps into an axial seed B-field to form co-rotating ion rings which merge near the mid-plane of the device and are subsequently radially compressed and azimuthally accelerated up to 1000 times the initial ion beam energies during the axial magnetic flux compression phase driven by the DPF plasma liner implosion. Two DPF and IRMG heads are implemented to dramatically reduce the size and cost of the drivers and increase ion ring capture efficiency along with the added benefits of favorable magnetic field line curvature throughout implosion process due to higher velocity shear-stabilized DPF plasma pinch flows near each gun muzzle and greater tuning capability for properly timing the implosion and ion beam generation, injection and compression of the two colliding and subsequently merged ion rings onto a solid or high energy density plasma target.

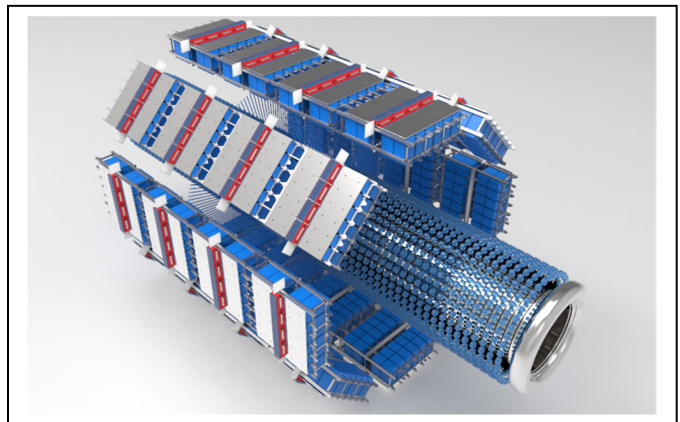

Fig. 1. Iso-view of The WARP Reactor

## III. STRATEGIC IMPORTANCE

WARP directly aligns with the DOE and NNSA missions and core competencies as an economically viable and climate-friendly rep-rated MagLIF device for nuclear fusion energy production as well as a multi-pulse compact flash x-ray/neutron source for assessing our aging nuclear weapons stockpile. In addition to developing the next-generation pulsed power architectures, this Strategic Initiative will help to benchmark present high-performance computing, simulation, and data science with respect to more disruptive and imaginative ultra-intense plasma/beam configurations. Finally, WARP success would fortify LLNL's place at the forefront of the subsequent RHED physics and technology revolution.

## IV. WARP REACTOR PHYSICS

The WARP Reactor conceptual design, models, simulations, and targeted performance characteristics for the various applications pull directly from standard fusion plasma, beam, accelerator and relativistic physics along with a modified Einstein Field Equation (EFE) [29] and respective Figures Of Merit (FOM) for assessing the validity of a recently proposed Naïve Quantum Gravity (NQG) theory [30].

### A. Ring Pinch and Acceleration Physics

The physics behind charged particle ring radial compression and azimuthal acceleration [31-40] in the WARP Core is as follows: Magnetic flux ($\Phi_z$) compression (i.e. conservation of seed $\Phi_z$ during DPF z-pinch driven imploding liners: $\Phi_z = B_z \pi r^2$ = constant) creates a "Magnetic Wave" (i.e. seed $B_z$ amplitude rapidly swells/increases since flux conservation dictates that $B_z \propto r^{-2}$) which forces (i.e. $F = qV \times B$) charged particle rings to radially compress (i.e. decrease in Larmor radius: $r_L = \gamma m V_\theta / qB$) and due to the conservation of canonical angular momentum (i.e. $p_\theta = \gamma m r V_\theta + q\Phi_z/2\pi$ = constant) and adiabatic flux conservation condition (i.e. $p_\theta^2 / B_z$ = constant or $p_\theta r$ = constant) the charged particle ring azimuthal velocities scale as: $V_\theta \propto (r_i/r_f)$ for $\gamma = 1$. Since $\Phi_z$ and $p_\theta$ are conserved in this device, final charged particle ring energy is: $E_f = \frac{1}{2} \gamma N m (V_{\theta f})^2 \sim E_i (r_i/r_f)^2$. Finally, for relativistic charged particle motion in pulsed B-fields, $\gamma$ varies due to $\nabla \times \mathbf{E} = -(\delta \mathbf{B}/\delta t)$ and therefore $E_f \sim E_i (B_f/B_i)$.

### B. Fusion Plasma Physics

The principal formulas used in the WARP Reactor conceptual models and simulations are the standard fusion plasma physics [41-45] equations in MKS units unless specifically identified otherwise (1)-(11) for: $N_n$ – the number of fusion-generated neutrons; $\beta$ – plasma to magnetic pressure; $E_f$ – total fusion energy; $E_p$ – plasma energy; $E_b$ – bremsstrahlung and $E_s$ – synchrotron radiation energy; $G_S$ – scientific and $G_E$ – engineering gains; $E_{sale}$ – fusion energy for sale; $v_A$ – Alfvén velocity; $\tau_R$ – magnetic reconnection time scales along with plasma and particle beam propagation modes (i.e. $\beta > 1$ for diamagnetic drift mode; $\beta \ll 1$ for collective mode; $\beta \ll 1$ with polarization E-field shorted for single particle mode).

$$N_n \sim n^2 <\sigma v> V \tau \tag{1}$$

$$\beta = \frac{n k T}{B^2/2\mu_0} \tag{2}$$

$$E_f = N_n E_r \tag{3}$$

$$E_p = \frac{3}{2} n V k T \tag{4}$$

$$Eb \sim 10^{-38} Z^2 n^2 T[ev]^{0.5} V \tau \tag{5}$$

$$E_s \sim \frac{2K q^2 \gamma^4 c}{3 r^2} N_e \tau \tag{6}$$

$$G_S = \frac{E_f}{E_p} \tag{7}$$

$$G_E = \frac{E_f}{E_T} \tag{8}$$

$$E_{sale} = f[E_f - E_p - E_b - E_s] \tag{9}$$

$$v_A = \frac{B}{\sqrt{\mu_0 \rho}} \tag{10}$$

$$\tau_R = \frac{L^2}{\delta v_A} \tag{11}$$

Where $n$ is the plasma density; $<\sigma v>$ is the fusion reaction rate; $V$ is the plasma/beam/ring volume; $\tau$ is the confinement time; $k$ is the Boltzmann constant; $T$ is plasma temperature; $B$ is the magnetic field; $\mu_0$ is the vacuum permeability; $E_r$ is the fusion energy per reaction; $Z$ is the atomic number; $K$ is the Coulomb constant; $q$ is the charge; $\gamma$ is the Lorentz factor; $c$ is the speed of light in vacuum; $N_e$ is the number of electrons; $E_T$ is the total stored energy of reactor; $f$ is the conversion efficiency; $\rho$ is the mass density; $L$ is the half-length of current sheet; and $\delta$ is the current sheet half-thickness.

### C. Relativistic Formulas, Modified EFE, NQG and FOM

In addition to the standard relativistic formulas for the Lorentz factor (12), momentum (13) and energy (14), we also introduce a modified EFE (15) with a NQG addition (16) that may be accessible by the WARP Reactor for verification or invalidation of the theory along with relevant FOM such as spacetime curvature (17), gravitational potential (18) and frame-dragging effects (19).

$$\gamma = \frac{1}{\sqrt{1-\frac{v^2}{c^2}}} \tag{12}$$

$$\vec{p} = \gamma m \vec{v} \tag{13}$$

$$E = \gamma m c^2 \tag{14}$$

$$G_{\mu\nu} = \frac{8\pi G}{c^4} (S + A)\, T_{\mu\nu} \tag{15}$$

$$T_{\mu\nu} \rightarrow Re[\frac{\psi_f^* \hat{T}_{\mu\nu} \psi_i}{<f|i>}] \quad (16)$$

$$C_{\alpha\sigma} = (S + A)\frac{G M}{c^2 V} \quad (17)$$

$$\Phi_{\alpha\sigma} = (S + A)\frac{G M}{c^2 R} \quad (18)$$

$$\Omega_{\alpha\sigma} = (S + A)\frac{G I \omega}{c^2 R^3} \quad (19)$$

$G_{\mu\nu}$ - Einstein curvature tensor; $T_{\mu\nu}$ - energy-momentum tensor; $8\pi G/c^4$ - energy-momentum to curvature coupling constant in vacuum; $G$ - Newton's gravitational constant; "S" - Sarfatti plasma metamaterial effects; "A" - Anderson Unruh/Casimir threshold effects; $T_{\mu\nu}$ - Sutherland NQG addition; $\hat{T}_{\mu\nu}$ - energy-momentum operator; $\psi_i$ and $\psi_f^*$ - initial and final conjugate wavefunctions, respectively; $<f|i>$ - final and initial boundary conditions; FOM: $C_{\alpha\sigma}$ - spacetime curvature; $\Phi_{\alpha\sigma}$ - gravitational potential and $\Omega_{\alpha\sigma}$ - frame-dragging effect; $M$ - ring mass; $R$ - ring radius; $I$ – ring moment of inertia; $\omega$ – ring angular velocity.

## V. WARP REACTOR TECHNOLOGY

The WARP Reactor utilizes tried-and-true Shiva Star-like "TEMPEST" Marx Modules to drive its dual Dense Plasma Focus head and state-of-the-art Impedance-matched Marx Generators (or Linear Transformer Drivers for rapid rep-rate operation) to drive the dual charged particle beam-ring reflex triodes. Figures 2, 3 and 4 show a side-view end-view and cross-sectional view of the full-scale WARP Reactor, respectively. The WARP Reactor consists primarily of 40 TEMPEST modules, 2 Ion Ring Marx Generators and the central WARP Core.

### A. TEMPEST Marx Modules

Forty TEMPEST Marx modules drive the dual DPF heads. A cross-sectional view of a single TEMPEST Marx module is provided in Figure 5 which consists of a more robust version of the Shiva Star design with upgraded 1.2MA railgap switches, a seismically-rated welded frame capacitor assembly along with a modified HV output header for the flexible high current coaxial cable connections. Each TEMPEST Module is primarily comprised of four super-duty railgap switches, aluminum parallel plate transmission lines and twenty-four +/-60kV, 250kA high energy density capacitors with a total energy stored per module of ~260kJ and a total 40 module TEMPEST system storage of >10MJ and capable of delivering ~60MA to the DPF loads.

### B. Ion Ring Marx Generators

Two IRMGs drive the dual Reflex Triode heads. For single pulse operation, each IRMG is a 30-stage Impedance-matched Marx Generator (IMG) [50] on steroids or a Linear Transformer Driver (LTD) [51] for rapid rate-rate operation. The IMG-version, cross-sectional view shown in Figure 6, consists primarily of 30 segmented coaxial return spools, a continuous tapered inner HV stalk, corona toroid, and 40 bricks per stage with each brick being comprised of two capacitors and a high-performance spark-gap switch. Each IRMG can produce a 1MV,

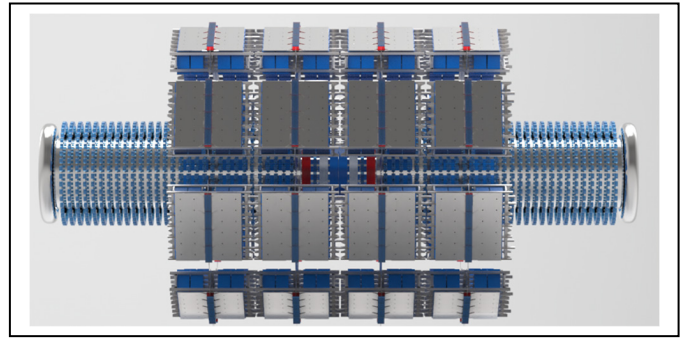

Fig. 2. Side-view of The WARP Reactor (total machine length ~ 18m)

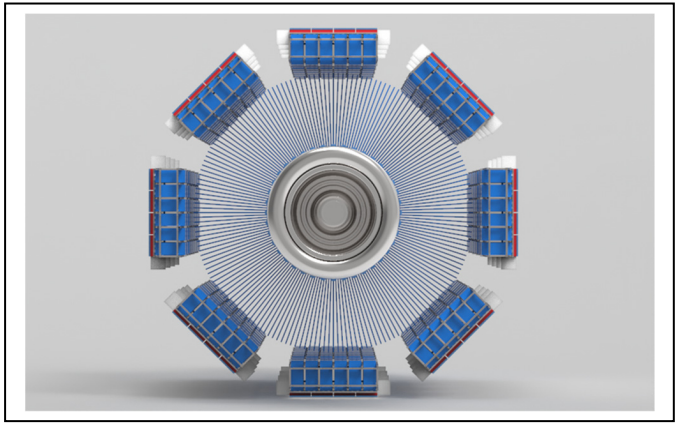

Fig. 3. End-view of The WARP Reactor (total machine diameter: ~ 9m)

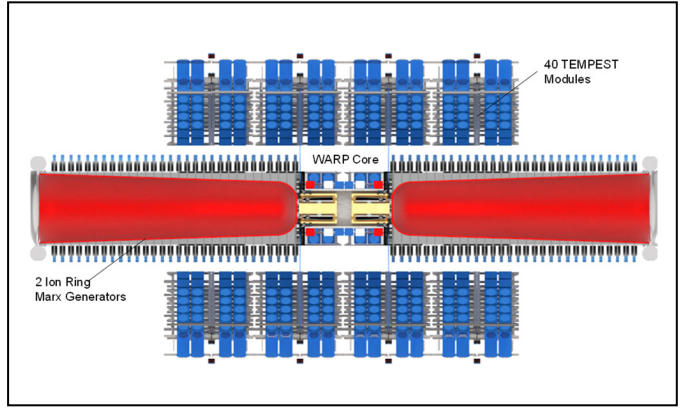

Fig. 4. Cross-sectional view of The WARP Reactor showing TEMPEST Modules, IRMGs and WARP Core

1MA, 100ns pulse to the Reflex Triode with a total energy storage for a two-IRMG system of ~2MJ.

### C. WARP Core

Depicted in Figure 7 is a cross-sectional view of the novel WARP Core (~1m L end-to-end x ~0.5m diameter at B-magnets). The relatively large Red, White and Blue squares situated around the grounded vacuum chamber represent the B-insulation, B-Seed and B-cusp pulsed magnets. Whereas the small blue squares (next to the magnets) and internal light blue rectangles are the annular DPF and IRMG puff valves, respectively. The TEMPEST Modules attach to the WARP Core at the DPF Collection Plates on the Left and Right End of the Device. Whereas the Ion Ring Marx Generators connect to the

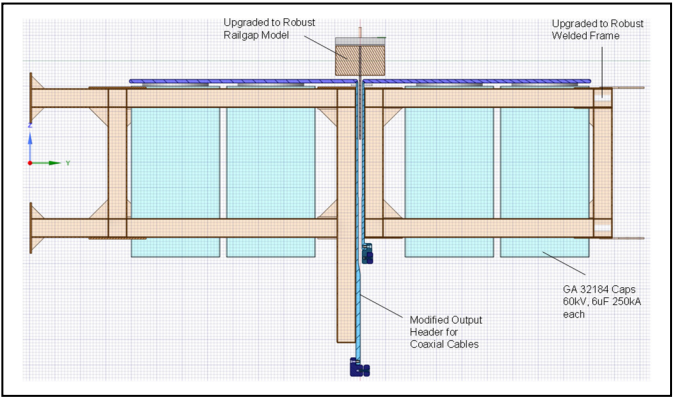

Fig. 5.  Cross-sectional view of TEMPEST Marx Module

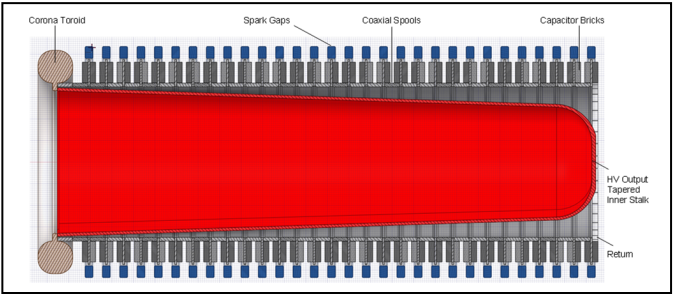

Fig. 6.  Cross-sectional view of Ion Ring Marx Generator (IMG-version)

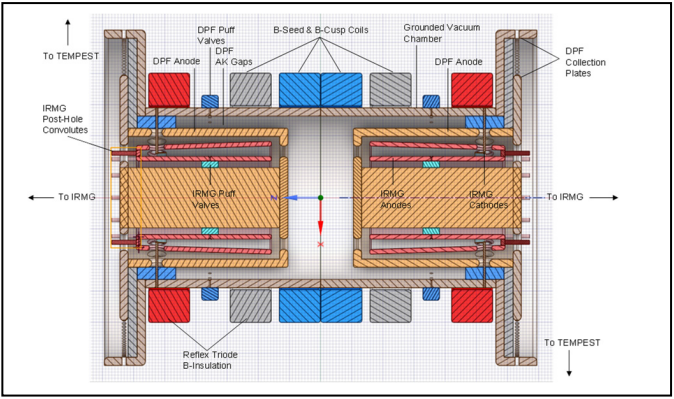

Fig. 7.  Cross-sectional view of the WARP Core

TABLE I.  WARP REACTOR MACHINE METRICS

| WARP-X (1/10 scale of Full Reactor): | 4 TEMPEST Modules | 2 IRMGs |
|---|---|---|
| Capacitor Specs | 60kV,6uF,250kA | 100kV,160nF,70kA |
| Total # of Caps | 96 | 64 |
| Railgap Model/Specs | 120kV, 1.2MA, 10C | 100kV, 120kA, 0.3C |
| Total Capacitance | 144uF erected | 160nF erected |
| Rated Voltage Reversal | 60% | 20% |
| Total # of Switches | 16 | 32 |
| Max Vcharge | +/-60kV | +/-50kV |
| Total Estored | 1.04MJ | 13kJ |
| Peak Vout | 100kV | 400kV |
| Peak Iout | 5MA | 400kA |
| Current Pulse Rise-time / Duration | 2.5us | 100ns |
| Rep-Rate | 1 shot/15 min | |
| WARP-X Dimensions | 20ft OD x 15ft H | |

| MACHINE Metrics & Comparisons: | PBFA II / Z | WARP-X | WARP-Reactor | WARP-R/PBFA II |
|---|---|---|---|---|
| DPF/Z-Pinch Current | NA / 26MA | 5MA | 60MA | 2.3x |
| Initial Ion Beam/Ring Current | 1MA | 0.2MA | 2MA | 2x |
| Final Ion Beam/Ring Current | 1MA | 2MA | 20MA | 20x |
| Initial Ion Beam/Ring Energy | 9MeV | 0.4MeV | 1MeV | 0.11x |
| Final Ion Beam/Ring Energy | 9MeV | 40MeV | 1GeV | 111x |
| Accel. Efficiency = $E_{beam}/E_{stored}$ | 0.66% | 15% | 25% | 38x |
| Machine Size | 108ft ODx16ft H | 20ft ODx15ft H | 30ft ODx55ft H | 0.96x |
| Total Cost Today | $250M | $6M | $150M | 0.6x |

## VI.  WARP REACTOR MACHINE METRICS

Table I shows the major pulsed power parameters and DPF plasma liner and Ion Beam/Ring metrics, respectively, along with order of magnitude comparisons with one of the largest ion beam accelerators in recent times, PBFA II. The WARP-X prototype is a 1/10-scale version of the full-scale Reactor which can deliver up to 5MA in < 3us into the DPF plasma liner whereas the 4-stage prototype version of the Ion Ring Marx Generators can generate initial ion beam energies and currents of up to 400keV with 200kA, respectively. The lower section of two back-ends of the dual Reflex Triodes through the IRMG Post-hole convolutes. Finally, the primary WARP Core and most novel central components consist of coaxial dual DPF (~13cm diameter) and IRMG heads with embedded Reflex Triodes (~10cm diameter).

Table I provides a comparison of plasma liner/beam/ion ring parameters between PBFA II/Z [52,53] and WARP devices with the 60MA WARP-R (Reactor) machine showing order of magnitude increases in ion beam energy (GeV-level) and current (20MA at implosion stagnation) with 25% acceleration efficiencies.

## VII.  WARP REACTOR OPERATIONS

Shown in Figures 8-12 are cross-sections of the WARP Core at different phases during operations: PHASE 1 begins with the creation of the B-insulation, B-seed and B-cusp magnetic fields to ensure full B-field diffusion through the metal structures along with gas injection via the DPF and IRMG annular puff valves and subsequent firing of the TEMPEST modules to initiate dual DPF plasmoid generation, lift-off and run-down. PHASE 2 occurs as the two DPFs begin their run-in sequence and the IRMGs are fired to produce the dual tubular ion beams. PHASE 3 is when the DPFs return currents have merged and the dual axially directed ion beams have passed through their respective B-cusp fields to become co-rotating and merged ion rings in the embedded B-seed field. PHASE 4 is the DPF plasma liner implosion sequence which provide the necessary flux compression and subsequent ion ring pinch and azimuthal acceleration. PHASE 5 is the implosion stagnation phase and relativistic high energy density ion ring and target interaction sequence for fusion energy production, advanced propulsion, super-flash x-ray and neutron generation for radiographic and/or dynamic interrogation applications and/or accessing new RHED physics regimes.

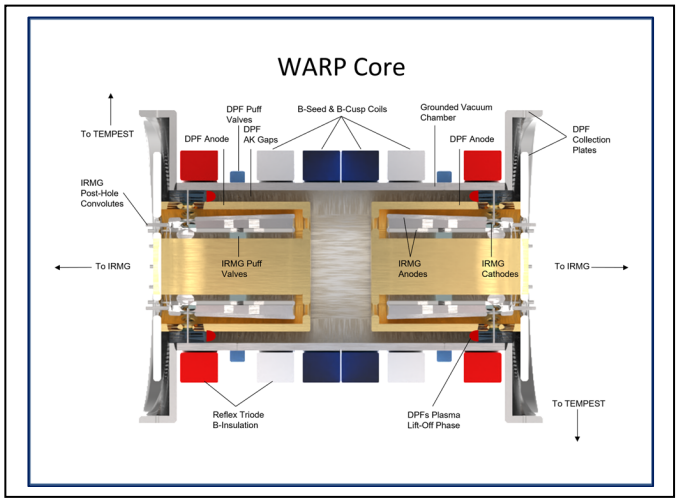

Fig. 8. WARP Reactor Operations: PHASE 1

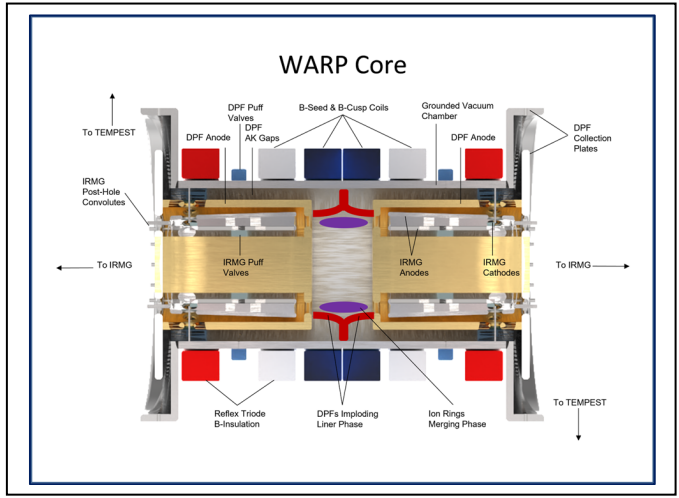

Fig. 11. WARP Reactor Operations: PHASE 4

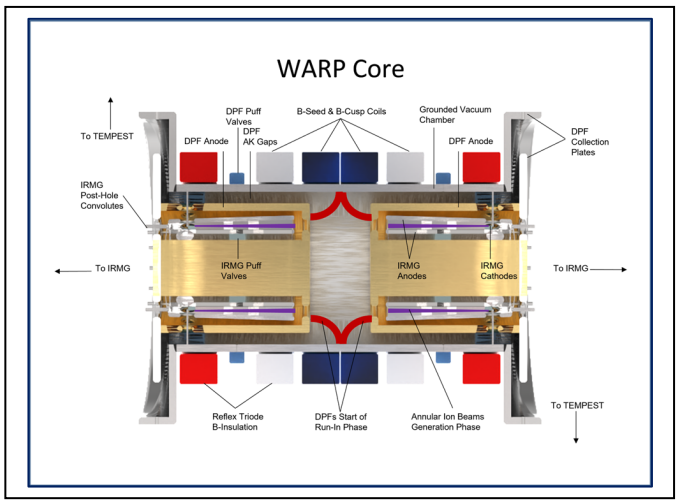

Fig. 9. WARP Reactor Operations: PHASE 2

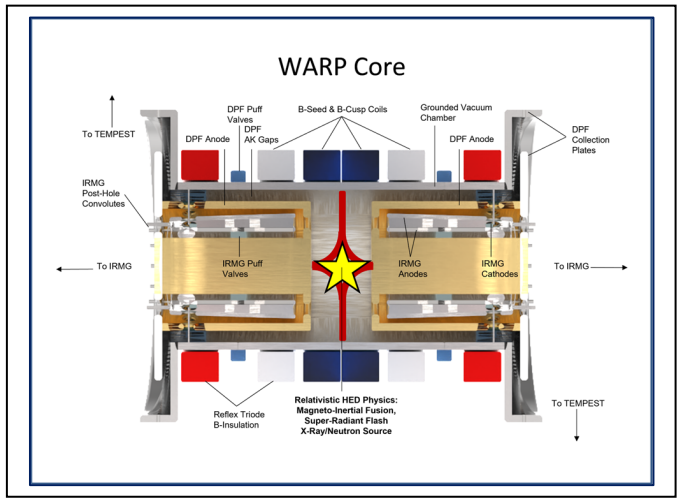

Fig. 12. WARP Reactor Operations: PHASE 5

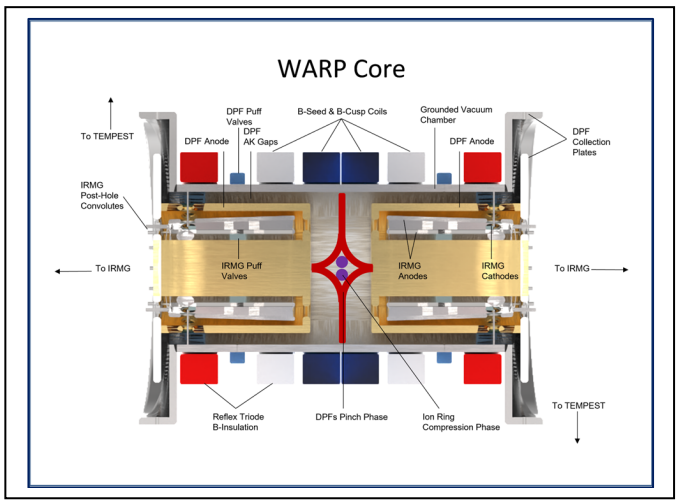

Fig. 10. WARP Reactor Operations: PHASE 3

To aid in visualizing the WARP Core Operations we have created a movie with respective phases identified. Figures 13-17 show the movie's progression from FRAME 1: the DPF plasma lift-off and run-down phases followed by FRAME 2: the Ion Beams Generation and DPF run-in phases then FRAME 3: the Ion Rings Generation and Merging phases followed by FRAME 4: the DPFs and Ion Ring Implosion and Acceleration phases and finally FRAME 5: the WARP Fusion phase, respectively.

## VIII. WARP REACTOR MODELS AND SIMULATIONS

To help benchmark our WARP Reactor's plasma liner implosion simulations we compare them to the Semi-Analytical Model for Magnetized Liner Inertial Fusion [54,55], outlined by Dr. McBride from University of Michigan and Dr. Slutz from Sandia, which can be used to reproduce the general 1D behavior of MagLIF machines [3-5]. This model provides many key aspects of MagLIF, including: (1) fuel preheat; (2) liner implosion; (3) liner compressibility, internal magnetic pressure, and ohmic heating; (4) adiabatic heating by compression; (5) fuel opacity and radiative loss; (6) B-flux compression with; (7) magnetized electron and ion thermal conduction losses; (8) end losses; (9) enhanced losses due to mixing; (10) D-D and D-T primary fusion reactions for fuel ratios; and (11) α-particle fuel heating. However, in order to expedite comparisons between our multi-physics models/simulations and semi-analytical models, we have created a simplified version of their MagLIF model in Mathematica which also allows rapid scanning of WARP Reactor input parameter space for obtaining desired operation scenarios and higher performance regimes.

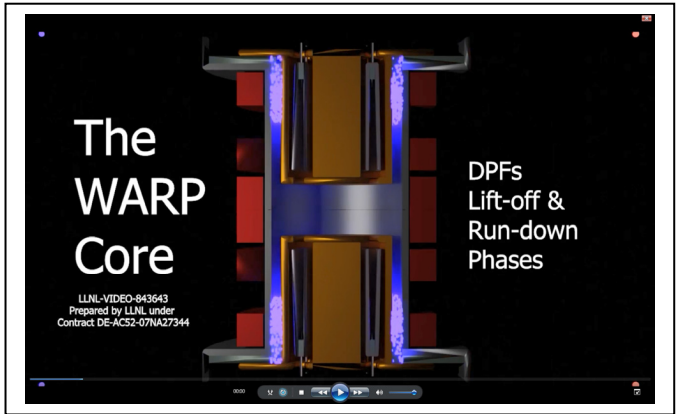

Fig. 13. WARP Reactor Movie Frame 1

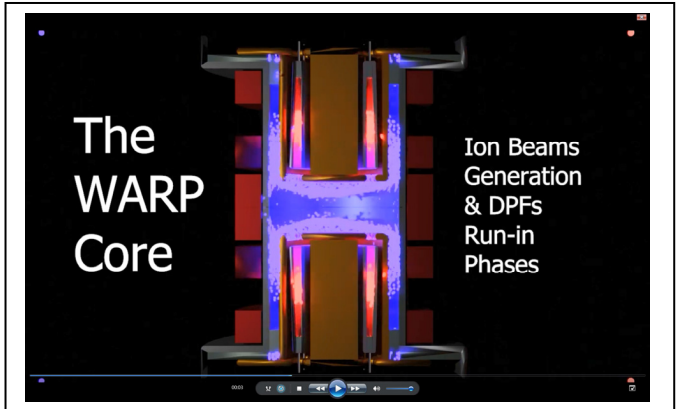

Fig. 14. WARP Reactor Movie Frame 2

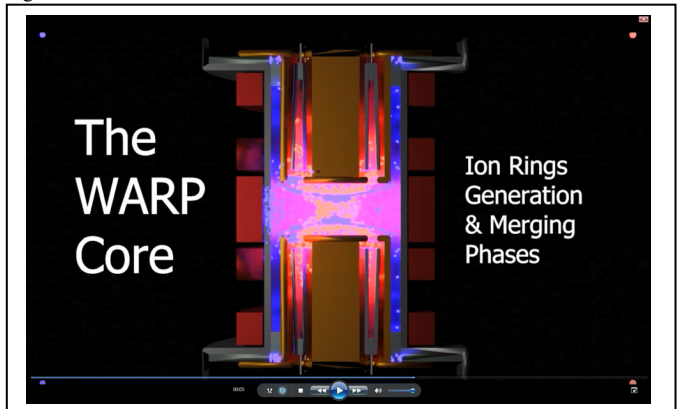

Fig. 15. WARP Reactor Movie Frame 3

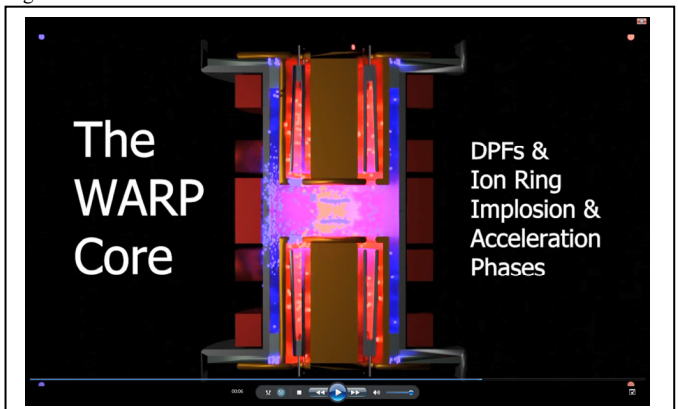

Fig. 16. WARP Reactor Movie Frame 4

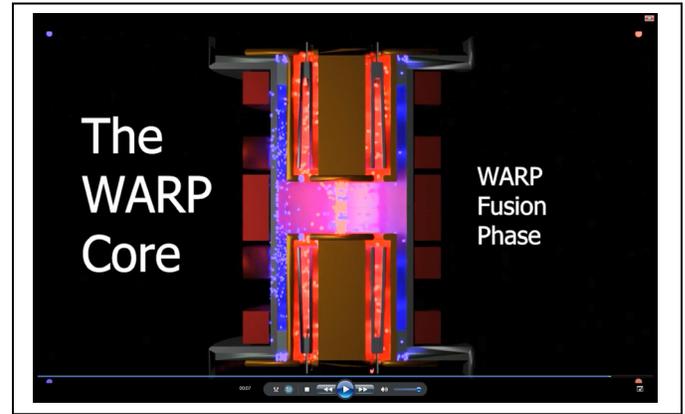

Fig. 17. WARP Reactor Movie Frame 5

In addition to the multi-physics models and simulations we of course have created 3D Computer Aided Design (CAD) engineering models with associated Finite Element Analysis (FEA) for electrical and mechanical stresses along with circuit models and simulations for the TEMPEST system, IRMGs (both for single-pulse IMG and rep-rated LTD varieties) and WARP Core dynamic loads, as shown in Figures 18 and 19, respectively.

Unfortunately, due to budget and computational constraints, a somewhat reduced simulation effort was performed for a mini-WARP Core in the Chicago particle-in-cell code in hybrid-kinetic mode where the electrons are treated as an inertia-less fluid using MHD equations for the electron response while the ions are treated as kinetic particles. Figure 20 is a graph of the R-Btheta (or radially enclosed current) and a graph of the respective plasma density at a specific time and both as a function of radial and axial position in centimeters. Figures 21-25 are snapshots of the simulation for the various phases from the dual DPF initial plasma generation, run-down, run-in to the merging and final implosion/pinch phases.

*A. DPF Plasmoids Magnetic Reconnection*

Table II captures the magnetic reconnection [56-57] time scales with respect to the DPF plasmoids characteristics as they propagate through their respective B-Cusp regions and just prior to their collision with one another at the midplane of the WARP Core.

*B. DPF & Ion Beam Propagation Mechanisms*

References [58-63] provide detailed propagation criteria for plasma/beam transport across magnetic field lines in vacuum or within a background magnetized plasma. Table III captures the

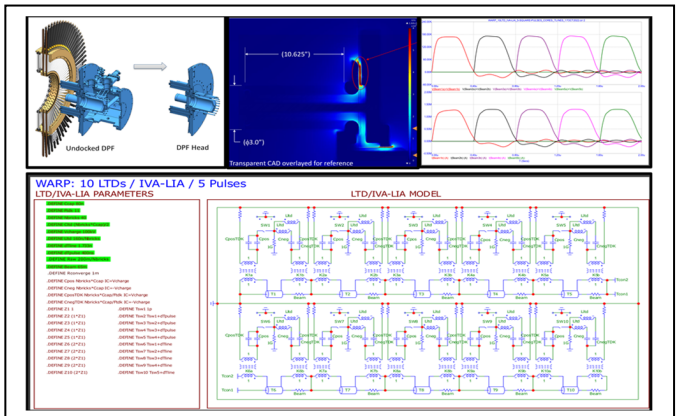

Fig. 18. WARP Reactor CAD modeling, FEA simulations, circuit models and analysis for multi-pulse IRMG (LTD-version) operations

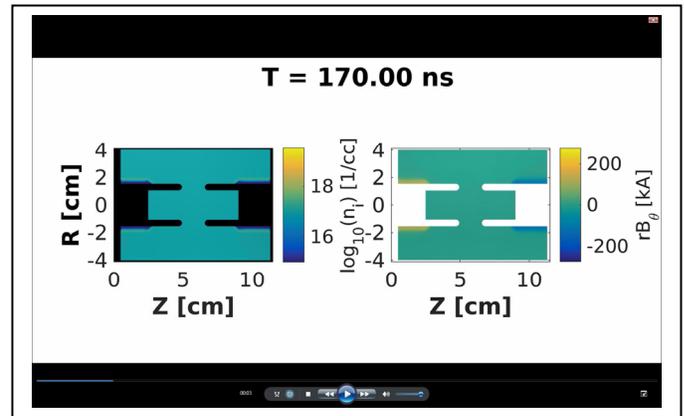

Fig. 21. Mini-WARP Core dual-DPF plasma generation phase

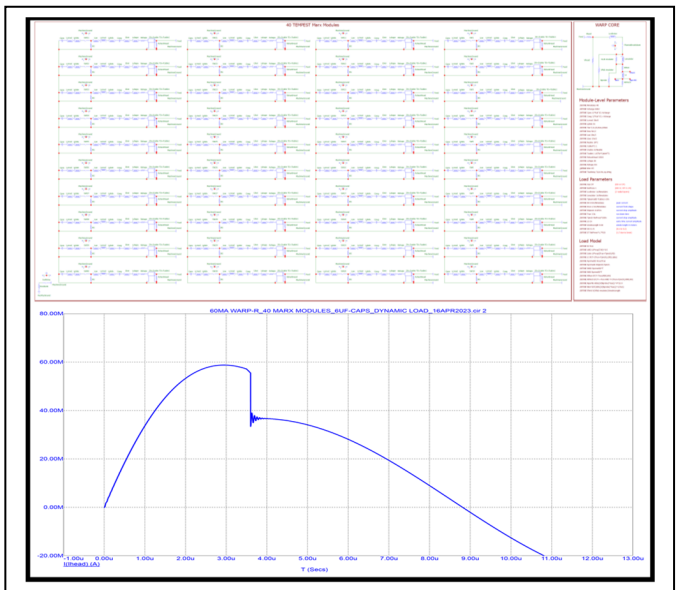

Fig. 19. 60MA WARP Reactor circuit model and analysis for 40 TEMPEST modules driving the WARP Core dynamic load

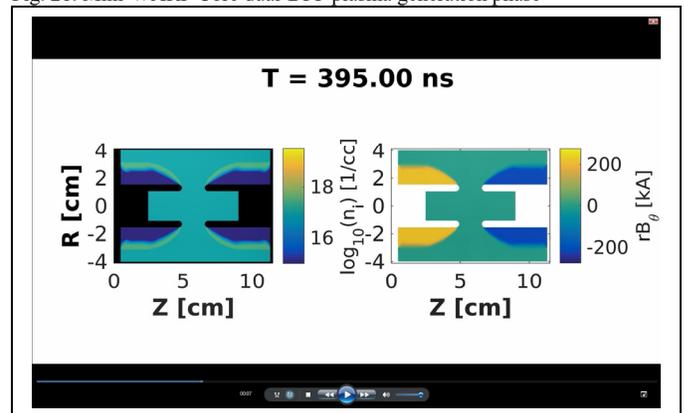

Fig. 22. Mini-WARP Core dual-DPF plasma run-down phase

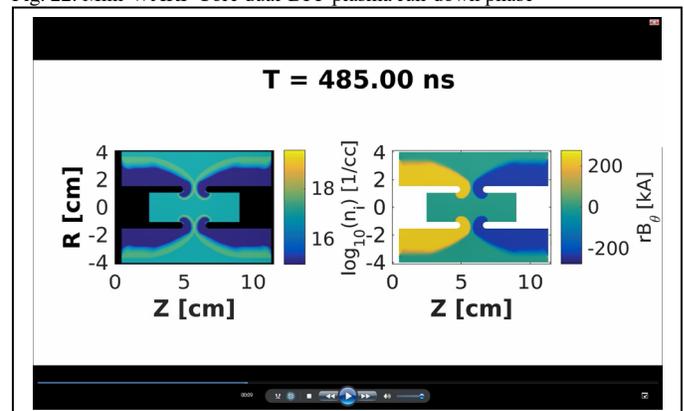

Fig. 23. Mini-WARP Core dual-DPF plasma run-in phase

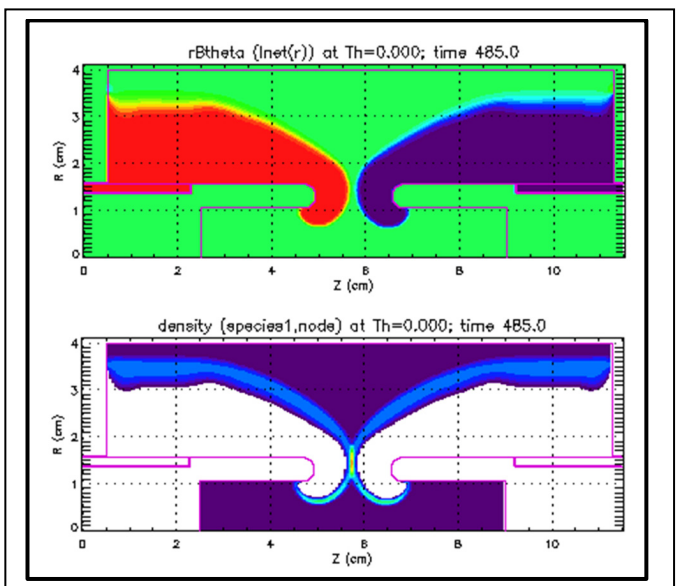

Fig. 20. Mini-WARP Core R-Btheta (top) and plasma density (bottom)

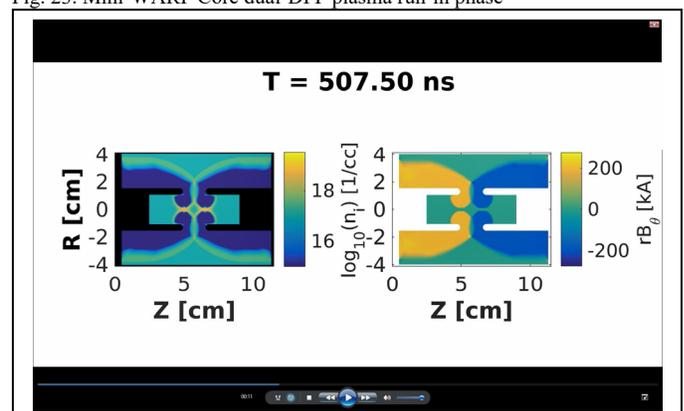

Fig. 24. Mini-WARP Core dual-DPF plasma merging phase

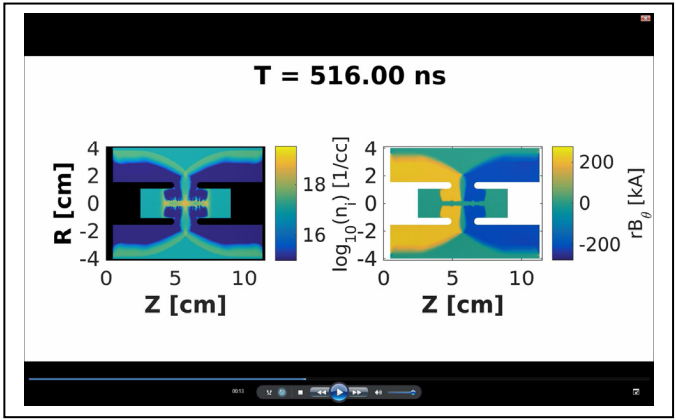

Fig. 25. Mini-WARP Core dual-DPF final implosion and pinch phases

DPF plasmoid initial characteristics and propagation mode via the diamagnetic drift mechanism for this parameter regime. Table IV shows the ion beam propagates in the collective mode unless the polarization E-field is shorted by background electrons. Finally, Table V displays the Reflex Triode and initial ion beam parameters along with final compression (millimeters scale) and accelerations (>$10^{19}$ m/s$^2$). NOTE: of particular interest to Warp, Unruh, dynamic Casimir and QG effects, the electron ring mode of operation indicates submicron implosion radii and >$10^{23}$ m/s$^2$ azimuthal accelerations depending on initial e-beam and magnetic flux parameters.

### C. Compression, Acceleration, Fusion & Radiation Metrics

Table VI captures the major ring compression and acceleration metrics while Table VII display the overall fusion and radiation metrics for DT operations, respectively. Finally, ion ring major radius at final implosion stagnation reaches the millimeters scale and at GeV-level azimuthal energies for protons, deuterons and tritons. With respect to the fusion/radiation parameters, the peak x-ray and DT fusion neutron yield reach 3.5MJ and 6.6x10^18 neutrons per pulse, respectively, with an overall scientific gain of >19.

### D. RHED Physics & Metric Engineering

In addition to the fusion energy and x-ray/neutron radiation applications, The WARP Reactor is also primed for accessing new RHED. The 60MA DPF plasma liner with embedded GeV-level ion ring appears capable of reaching thousands of Tesla and subsequently multi-TPa pressures as captured in Table VIII. Due to the dramatic compression and acceleration of the charged particle rings and plasma liner/target interactions we may even begin to generate significant plasma metamaterial, Warp, Unruh and Dynamic Casimir effects not to mention copious amounts of synchrotron radiation from electron ring modes of operations. Furthermore, once the charged particle rings become relativistic during the final moments of implosion phase the decrease in ring radius saturates as relativistic mass increases. In other words, flux compression shifts from ring velocity to ring mass or enhanced spacetime coupling. Finally, the WARP Reactor may also provide a means for accessing the Quantum Gravity (QG) domain for verification or invalidation of various proposed theories by probing for any spacetime curvature, gravitational potential and/or frame-dragging effects. With respect to our modified EFE and FOM equations (15)-(19) and Mathematica models (see Figure 26), we find a significant enhancement of

TABLE II. DPF PROPAGATION AND MAGNETIC RECONNECTION TIMES

| Parameter | Value | Unit |
|---|---|---|
| DPF plasma current length (L) | 0.03 | m |
| DPF plasma current layer thickness ($\delta$) | 0.03 | m |
| Magnetic B-cusp field (B) | 1 | T |
| Deuteron mass ($m_d$) | 3.34x10$^{-27}$ | kg |
| Plasma ion density (n) | 2x10$^{23}$ | m$^{-3}$ |
| Plasma ion mass density ($\rho$) | 6.68x10$^4$ | kg m$^{-3}$ |
| Alfven velocity ($V_A$) | 3.45x10$^4$ | m/s |
| Magnetic reconnection time ($T_R$) | 8.69x10$^{-7}$ | s |
| Plasmoid velocity ($V_{plasmoid}$) | 10$^5$ | m/s |
| Distance to collision with other plasmoid (D) | 0.10 | m |
| Time to collision after passing through B-cusp ($T_C$) | 1.02x10$^{-6}$ | s |
| Magnetic reconnection to plasmoid collision times ($T_R/T_C$) | 0.85 | |

TABLE III. DPF PLASMOID PROPAGATION MODE

| Diamagnetic Propagation when $nkT/(B^2/2\mu_0) = \beta > 1$ | | |
|---|---|---|
| Parameter | Value | Unit |
| Initial plasma density ($n_i$) | 2x10$^{23}$ | m$^{-3}$ |
| Plasma temperature (T) | 15 | eV |
| Plasma pressure (nkT) | 4.8x10$^5$ | Pa |
| Cusp magnetic field (B) | 1 | T |
| Magnetic pressure ($B^2/2\mu_0$) | 3.98x10$^5$ | Pa |
| Plasma/magnetic pressure ($\beta$) | 1.2 | satisfied |

TABLE IV. ION BEAM PROPAGATION MODES

| Collective mode of propagation when $nkT/(B^2/2\mu_0) = \beta \ll 1$ and . . . | | |
|---|---|---|
| Parameter | Value | Unit / Comment |
| Initial beam density ($n_i$) | 4.1x10$^{18}$ | m$^{-3}$ |
| Ion thermal ($T_i$) | 1000 | eV |
| Electron thermal ($T_e$) | 500 | eV |
| Ion beam directed energy ($E_{ib}$) | 4.0x10$^5$ | eV |
| Beam thermal pressure (nkT) | 658 | Pa |
| Initial magnetic field ($B_i$) | 1 | T |
| Magnetic pressure ($B^2/2\mu_0$) | 3.98x10$^5$ | Pa |
| Beam thermal/magnetic pressure condition ($\beta \ll 1$) | 1.65x10$^{-3}$ | satisfied |
| Square root of mass ratio $(m_i/m_e)^{1/2}$ | 60.6 | |
| Deuterium ion mass ($m_i$) | 3.34x10$^{-27}$ | kg |
| Electron mass ($m_e$) | 9.1x10$^{-31}$ | kg |
| Beam ion mass density ($\rho$) | 1.37x10$^{-8}$ | kg m$^{-3}$ |
| Alfven velocity ($V_A$) | 7.62x10$^6$ | m/s |
| Vacuum permittivity ($\varepsilon_0$) | 8.85x10$^{-12}$ | F/m |
| Beam dielectric constant ($\varepsilon$) | 1.37x10$^{-8}$ | F/m |
| Beam relative permittivity ($\varepsilon_R$) | 1551 | |
| E-field/energy density condition ($\varepsilon_R/(m_i/m_e)^{1/2} \gg 1$) | 25.6 | satisfied |
| Beam directed flow ($V_{ib}$) | 6.19x10$^6$ | m/s |
| Beam directed pressure ($nmv^2/2$) | 2.63x10$^5$ | Pa |
| Beam directed/magnetic pressure condition ($\beta \ll 1$) | 0.66 | marginally satisfied |
| Ions thermal velocity ($V_{thermal(ion)}$) | 3.1x10$^5$ | m/s |
| Electrons thermal velocity ($V_{thermal(e)}$) | 1.33x10$^7$ | m/s |
| Beam radius ($R_b$) | 0.02 | m |
| Beam initial Larmor radius ($R_{Lib}$) | 0.13 | m |
| Thermal ion Larmor raidus ($R_{Li}$) | 0.0065 | m |
| Thermal electron Larmor radius ($R_{Le}$) | 7.54x10$^{-5}$ | m |
| Beam polarized sheath thickness ($\delta$) | 4.16x10$^{-6}$ | m |
| Polarization condition ($\delta/R_b \ll 1$) | 0.0002 | satisfied |
| Beamlet condition ($4R_b/R_L^i < 1$) | 12.38 | not satisfied (if E-field not shorted) |
| Plasma frequency ($\omega_i$) | 1.89x10$^9$ | rad/s |
| Ion cyclotron frequency ($\Omega_i$) | 4.79x10$^7$ | rad/s |
| Electron cyclotron frequency ($\Omega_e$) | 1.76x10$^{11}$ | rad/s |
| Virtual anode formation distance (L) | 0.002 | m |
| Virtual anode condition (($R_L/\varepsilon$)/L $\ll 1$) | 0.0019 | satisfied |

TABLE V. Reflex Triode and Beam Parameters

| Parameter | Value | Unit |
|---|---|---|
| Anode length | 0.127 | m |
| Ions per unit length | $10^{17}$ | #/m |
| Number of ions | $1.27 \times 10^{16}$ | # |
| Length of beam | 0.095 | m |
| Radius of beam | 0.02 | m |
| Area of beam | 0.032 | $m^2$ |
| Volume of beam | 0.003 | $m^3$ |
| Density of beam | $4.1 \times 10^{18}$ | $m^{-3}$ |
| Single ion beam current | $1.3 \times 10^5$ | A |
| Two ion beams current | $2.6 \times 10^5$ | A |
| Erected Marx voltage | $4.0 \times 10^5$ | V |
| Triode impedance | 1.57 | Ohms |
| Single IRMG current | $2.5 \times 10^5$ | A |
| Two IRMGs current | $5.0 \times 10^5$ | A |
| Ion beam/IRMG efficiency | 51.7 | % |

| Parameter | Ion Ring Value | Electron Ring Value | Unit |
|---|---|---|---|
| Final magnetic field | 1250 | 1250 | T |
| Final ring major radius | 0.005 | $2.73 \times 10^{-6}$ | m |
| Final ring volume | $1.55 \times 10^{-9}$ | $4.6 \times 10^{-16}$ | $m^3$ |
| Final ring density | $1.64 \times 10^{25}$ | $5.51 \times 10^{31}$ | $m^{-3}$ |
| Particle acceleration | $1.8 \times 10^{19}$ | $3.3 \times 10^{22}$ | $m/s^2$ |
| Final magnetic field | 5000 | 5000 | T |
| Final ring major radius | 0.003 | $3.4 \times 10^{-7}$ | m |
| Final ring volume | $6.06 \times 10^{-10}$ | $7.20 \times 10^{-18}$ | $m^3$ |
| Final ring density | $4.19 \times 10^{25}$ | $1.14 \times 10^{36}$ | $m^{-3}$ |
| Particle acceleration | $2.87 \times 10^{19}$ | $2.64 \times 10^{23}$ | $m/s^2$ |

TABLE VI. Ring Compression and Acceleration Metrics

| Parameter | Value | Unit |
|---|---|---|
| Speed of light in vacuum | 299939418 | m/s |
| Elementary charge | $1.6 \times 10^{-19}$ | C |
| Proton mass | $1.67 \times 10^{-27}$ | kg |
| Deuteron mass | $3.34 \times 10^{-27}$ | kg |
| Triton mass | $5.01 \times 10^{-27}$ | kg |
| Initial ring energy | $10^6$ | eV |
| Final ring energy | $3.9 \times 10^9$ | eV |
| Final to initial ring energy | 3906.25 | |
| Initial ring radius | 0.25 | m |
| Final ring radius | 0.004 | m |
| Initial to final ring radius | 62.5 | |
| Initial magnetic field | 1 | T |
| Final magnetic field | 3906.25 | T |
| Final to initial magnetic field | 3906.25 | |
| Proton initial velocity | 13848418 | m/s |
| Deuteron initial velocity | 9792310 | m/s |
| Triton initial velocity | 7995387 | m/s |
| Initial proton Lorentz factor | 1.00106 | |
| Initial deuteron Lorentz factor | 1.00053 | |
| Initial triton Lorentz factor | 1.00035 | |
| Final proton Lorentz factor | 5.16 | |
| Final deuteron Lorentz factor | 3.08 | |
| Final triton Lorentz factor | 2.38 | |
| Proton final azimuthal velocity | 291144639 | m/s |
| Deuteron final azimuthal velocity | 263001597 | m/s |
| Triton final azimuthal velocity | 207792135 | m/s |
| Proton ring final Larmor radius | 0.00401 | m |
| Deuteron ring final Larmor radius | 0.00433 | m |
| Triton ring final Larmor radius | 0.00397 | m |
| Proton ring final azimuthal energy | $3.9 \times 10^9$ | eV |
| Deuteron ring final azimuthal energy | $3.9 \times 10^9$ | eV |
| Triton ring final azimuthal energy | $3.9 \times 10^9$ | eV |

energy-momentum to spacetime curvature coupling due to plasma metamaterial effects via the Sarfatti "S" field factor; the initiation of direct spacetime metric phase change generated by accelerating and imploding RHED charged particle rings beyond Unruh/Casimir thresholds via the Anderson "A" field factor and QG effects via Sutherland's NQG theory with predicted major results provided in Table IX.

## IX. Conclusion

In conclusion, we envision the WARP Reactor as a more compact, modular and economically viable magneto-inertial fusion device for energy production (i.e. G > 19/pulse, nTτ ~ $5.2 \times 10^{21}$ keVs/$m^3$) or advanced propulsion and/or a super-radiant flash x-ray/neutron source for dynamic radiographic applications (i.e. x-ray/neutron yields per pulse > 3.5MJ/$6.6 \times 10^{18}$, Avg. Luminance ~ $10^{25}$ x-ray photons / s $mm^2$ $mrad^2$) along with providing access to new relativistic high energy density physics regimes (i.e. multi-MA, GeV-level charged plasma/particle beam-target interactions at multi-TPa, Unruh, Dynamic Casimir & QG effects). The WARP devices would fulfill the immediate need for an intermediate-level machine for z-pinch, beam and pulsed-power flow studies along with the added benefit of recruiting the next-generation of RHED plasma and accelerator scientists, engineers and technicians. Finally, WARP would be an ideal platform for prototyping novel pulsed power architectures for continuous rep-rate nuclear fusion and radiographic movie operations thereby enabling us to continue our collaborations across the Department of Energy (DOE)/Department of Defense (DOD) complexes along with forging new university and private industry partners through our Cooperative Research and Development Agreement (CRADA) and Strategic Partnership (SPP) programs.

We leave you with the following Gedankenexperiment we call "A Twisted Compression of the Ehrenfest Paradox" along with three conjectures on how one might enhance energy-momentum to spacetime curvature coupling. Imagine you are one of a multitude of elementary charged particles which make up a rotating high-energy-density charge/current neutralized particle ring, embedded within an axial seed magnetic field, that is radially compressed toward zero radius and azimuthally accelerated to ultra-relativistic velocity during sufficient flux compression. According to you and a nearby inertial observer, what happens to the local spacetime surrounding said ring-vortex at the end of the implosion phase?

TABLE VII.    FUSION AND RADIATOIN METRICS FOR DT OPERATIONS

| Parameter | Value | Unit |
|---|---|---|
| Total machine energy stored ($E_o$) | $10^7$ | J |
| Final magnetic field ($B_f$) | 5000 | T |
| Final plasma temperature ($T_f$) | 50000 | eV |
| Confinement time ($\tau$) | $10^{-7}$ | s |
| Final plasma density ($n_f$) | $1.04 \times 10^{27}$ | $m^{-3}$ |
| Final plasma volume ($Vol_f$) | $7.85 \times 10^{-8}$ | $m^3$ |
| Number of ions (N) | $8.13 \times 10^{19}$ | # |
| Average ion velocity ($<v>$) | $1.96 \times 10^6$ | m/s |
| DT fusion cross section ($<\sigma>_{DT}$) | $4.0 \times 10^{-28}$ | $m^2$ |
| DT reaction rate ($<\sigma v>_{DT}$) | $7.83 \times 10^{-22}$ | $m^3$/s |
| # of fusion neutrons per pulse ($N_n$) | $6.6 \times 10^{18}$ | #/pulse |
| Bremsstrahlung energy per pulse ($E_{brem}$) | $3.18 \times 10^4$ | J/pulse |
| Synchrotron energy per pulse ($E_{sync}$) | $3.21 \times 10^6$ | J/pulse |
| Fusion energy per pulse ($E_f$) | $1.86 \times 10^7$ | J/pulse |
| Plasma energy per pulse ($E_p$) | $9.77 \times 10^5$ | J/pulse |
| Scientific gain ($E_f/E_p$) | 19 | fraction |
| Energy for sale ($E_{sale}$) | $7.17 \times 10^6$ | J/pulse |
| Fusion double product ($n\tau$) | $1.04 \times 10^{20}$ | s/$m^3$ |
| Fusion triple product ($nT\tau$) | $5.18 \times 10^{21}$ | keV s/$m^3$ |
| $n\tau$/$Lawson_{DT}$ | 0.69 | |
| $nT\tau$/$LawsonTriple_{DT}$ | 1.73 | |

TABLE VIII.    RELATIVISTIC HIGH ENERGY DENSITY METRICS

| Parameter | DPF Plasma Liner Value | Unit |
|---|---|---|
| Final plasma density (n) | $10^{27}$ | $m^{-3}$ |
| Plasma temperature (T) | 50000 | eV |
| Final plasma pressure (nkT) | $8.0 \times 10^{12}$ | Pa |
| Final magnetic field (B) | 5000 | T |
| Magnetic pressure ($B^2/2\mu_0$) | $9.95 \times 10^{12}$ | Pa |
| Plasma/magnetic pressure ($\beta$) | 0.8 | |

| Parameter | Ion Ring Value | e-Ring Value | Unit |
|---|---|---|---|
| Final magnetic field | 5000 | 5000 | T |
| Final ring major radius | 0.003 | $3.4 \times 10^{-7}$ | m |
| Final ring volume | $6.06 \times 10^{-10}$ | $7.20 \times 10^{-18}$ | $m^3$ |
| Final ring density | $4.19 \times 10^{25}$ | $1.14 \times 10^{36}$ | $m^{-3}$ |
| Particle acceleration | $2.87 \times 10^{19}$ | $2.64 \times 10^{23}$ | m/$s^2$ |

```
μp = 1 ;  (*plasma relative permeability [dimensionless]*)
ωp = √(n q²/(ε0 me)) ;  (*plasma frequency [rad/s]*)
ω = 2π/Tpulse ;  (*EM pulse drive frequency [rad/s]*)
εp = 1 + ωp²/ω² ;  (*plasma relative permittivity*)
ar = (ae/aθ)² ;  (*electron ring relative acceleration to Unruh effect threshold [dimensionless]*)
Rr = (Rering/R0)³ ;  (*electron ring relative radius to Casimir effect threshold [dimensionless]*)
Sfield = ½ (εp² + 1/μp²) ;  (*Sarfatti S-field [dimensionless]*)
Afield = (ar/Rr)² ;  (* Anderson A-field [dimensionless]*)
σGμν = (8πG/c⁰⁴) Sfield Tμν ;  (*Sarfatti-only spacetime curvature [m⁻²]*)
αGμν = (8πG/c⁰⁴) Afield Tμν ;  (*Anderson-only spacetime curvature [m⁻²]*)
ασGμν = (8πG/c⁰⁴) (Afield + Sfield) Tμν ;  (*Anderson-Sarfatti spacetime curvature [m⁻²]*)
Cασ = (Afield + Sfield) G Mring/(c0² Vring) ;  (*Anderson-Sarfatti spacetime curvature approximation [m⁻²]*)
φασ = (Afield + Sfield) G Mring/(c0² Rring) ;  (*Anderson-Sarfatti gravitational potential approximation [dimensionless]*)
Ωασ = (Afield + Sfield) ((G Ir ωr)/(c0² RL³)) ;
(*Figure of Merit: Anderson-Sarfatti Lense-Thirring internal angular velocity of inertial dragging field [rad/s]*)
```

Fig. 26. Mathematica models for proposed modified EFE and FOM with Sarfatti and Anderson fields

TABLE IX.    METRIC ENGINEERING

| Parameter | Value | Unit |
|---|---|---|
| Particle density ($n_{i,e}$) | $1.047 \times 10^{27}$ | $m^{-3}$ |
| Elementary charge (q) | $1.6 \times 10^{-19}$ | C |
| Electron mass ($m_e$) | $9.1 \times 10^{-31}$ | kg |
| Plasma frequency ($\omega_{pe}$) | $1.82 \times 10^{15}$ | rad/s |
| proton mass ($m_p$) | $1.67 \times 10^{-27}$ | kg |
| EM-pulse drive frequency ($\omega = 2\pi/T$) | $1.57 \times 10^7$ | rad/s |
| Non-magnetized plasma relative permittivity ($\varepsilon_r$) | $-1.35 \times 10^{16}$ | |
| Magnetized plasma relative permittivity ($\varepsilon_{mr}$) | 7910 | |
| Spacetime curvature coupling in vacuum ($8\pi G/c^4$) | $2.07 \times 10^{-43}$ | 1/N |
| Plasma refractive index (n) | $1.16 \times 10^8$ | |
| Sarfatti scalar field factor (S) | $9.085 \times 10^{31}$ | |
| Anderson scaler field factor (A) | $1.9 \times 10^{30}$ | |
| Einstein tensor to stress-energy tensor ($G_{\mu\nu}/T_{\mu\nu}$) | $1.89 \times 10^{-11}$ | 1/N |
| Energy-momentum tensor --> energy density ($T_{\mu\nu}$) | $1.57 \times 10^{17}$ | N/$m^2$ |
| Einstein tensor --> spacetime curvature ($G_{\mu\nu}$) | $2.97 \times 10^6$ | 1/$m^2$ |
| Reciprocal of Einstein tensor ($1/G_{\mu\nu}$) | $3.37 \times 10^{-7}$ | $m^2$ |
| Final area of particle ring ($A_{ring}$) | $1.23 \times 10^{-6}$ | $m^2$ |
| Figure of merit for spacetime curvature ($C_{\alpha\sigma}$) | $8.64 \times 10^4$ | 1/$m^2$ |
| Figure of merit for gravitational energy ($\phi_{\alpha\sigma}$) | $3.35 \times 10^{-3}$ | |
| Figure of merit for frame-dragging effect ($\Omega_{\alpha\sigma}$) | $2.3 \times 10^{10}$ | rad/s |
| Warp factor ($A_{ring}/(1/G_{\mu\nu})$) | 3.66 | |

Finally, the three possible methods for enhancing energy-momentum to spacetime curvature coupling are as follows:

1. Multi-layer RHED plasma and charged particle ring confinement of THz radiation to create plasma/ring metamaterial effects.

2. Azimuthal acceleration beyond Unruh threshold of multi-pass RHED plasma and charged particle rings to generate Leidenfrost-like vortex layers which create spacetime phase transition.

3. Implosion beyond Casimir threshold of RHED plasma and charged particle rings to generate internal negative energy density which also provides additional confinement mechanism to the >5kT magnetic fields in order to prevent plasma/ring metamaterial rapid disassembly.

### ACKNOWLEDGMENT

The authors would very much like to thank Keith LeChien, Stephen Sampayan and Nathan Meezan for collaboration, reviews, critical discussions and support. This work was performed under the auspices of the U.S. Department of Energy by Lawrence Livermore National Laboratory under Contract DE-AC52-07NA27344.